\documentclass{ifacconf}

\usepackage{natbib}        
\usepackage{xcolor}
\usepackage{url}
\usepackage{multirow} 
\usepackage{array} 
\usepackage{booktabs}
\usepackage{threeparttable}
\usepackage{arydshln}  
\usepackage{graphicx}
\usepackage{subfigure}     
\usepackage{makecell}
\begin{document}
\begin{frontmatter}

\title{The Lazy Student's Dream: ChatGPT Passing an Engineering Course on Its Own} 

\author{Gokul Puthumanaillam, Timothy Bretl, Melkior Ornik}

\address{University of Illinois Urbana-Champaign, Urbana, IL, USA \\
(e-mails: \texttt{\{gokulp2, tbretl, mornik\}@illinois.edu})}
\begin{abstract}                
This paper presents a comprehensive investigation into the capability of Large Language Models (LLMs) to successfully complete a semester-long undergraduate control systems course. Through evaluation of 115 course deliverables, we assess LLM performance using ChatGPT under a ``minimal effort" protocol that simulates realistic student usage patterns. The investigation employs a rigorous testing methodology across multiple assessment formats, from auto-graded multiple choice questions to complex Python programming tasks and long-form analytical writing. Our analysis provides quantitative insights into AI's strengths and limitations in handling mathematical formulations, coding challenges, and theoretical concepts in control systems engineering. The LLM achieved a B-grade performance (82.24\%), approaching but not exceeding the class average (84.99\%), with strongest results in structured assignments and greatest limitations in open-ended projects. The findings inform discussions about course design adaptation in response to AI advancement, moving beyond simple prohibition towards thoughtful integration of these tools in engineering education. Additional materials including syllabus, examination papers, design projects, and example responses can be found at the project website: 
\textcolor{teal}{\url{https://gradegpt.github.io}}.

\end{abstract}


\end{frontmatter}

\section{Introduction}
The rapid emergence of Large Language Models (LLMs) in educational settings \citep{a2} has sparked substantial discussion about their impact on academic integrity and learning outcomes. While numerous studies have examined LLMs' capabilities in specific tasks \citep{a5} or isolated assignments \citep{a3}, there remains a critical gap in our understanding of their performance across an entire technical course. This work presents a systematic investigation of LLM performance across a complete undergraduate Aerospace Control Systems course (AE 353) at the University of Illinois Urbana-Champaign (UIUC), with the goal of quantifying the LLM's capabilities in handling diverse engineering coursework with no expert interaction and informing future course design strategies.

The remarkable versatility of LLMs in solving diverse problems, from creative writing to technical analysis, has led to widespread adoption among students, raising legitimate concerns about assessment integrity \citep{a6} and educational effectiveness. Studies across various disciplines have demonstrated these models' ability to handle academic tasks ranging from essay composition to mathematical problem-solving. Yet, their effectiveness in comprehensive technical courses, particularly those requiring theoretical and practical knowledge, remains unexplored. This gap in understanding is especially relevant in control systems education, where success demands not just isolated problem-solving but a coherent grasp of interconnected concepts across theory and application.

Our study examines AE 353, a junior-level control systems course, evaluating LLM performance across approximately 115 course deliverables. The course materials span multiple assessment formats: auto-graded homework problems, written examinations requiring mathematical derivations, and extensive programming projects. In investigating LLM capabilities, we chose to use ChatGPT, the most widely accessible and commonly used LLM among students. Our approach simulates a realistic scenario of a student seeking to complete coursework with minimal investment of time and effort---simply copying and pasting questions into ChatGPT without providing additional guidance.

\textit{Statement of contributions: }
Our research deliberately maintains a focused scope, examining LLM performance in a single undergraduate control systems course rather than attempting to draw broad conclusions about AI in education. The significance of this work lies in two key areas: First, we establish a rigorous, reproducible methodology for evaluating LLM performance across diverse assessment types within technical courses. Second, we provide quantitative insights into the current capabilities and limitations of LLMs in completing complex engineering coursework.

This study emphasizes technical performance evaluation, intentionally setting aside broader questions about educational policy and ethics of AI use in academia. While these questions merit thorough investigation, our priority is to establish a factual foundation regarding LLM capabilities in a specific technical course. While our findings are specific to AE 353 during Fall 2024, the analytical approach contribute to the broader dialogue about engineering education in an era of readily available AI tools.

\section{Background and Related Work}
%
\subsection{Current State of LLMs in Education}
The educational impact of LLMs has been extensively studied across multiple dimensions since their widespread availability. Early research by \citep{a7} documented the rapid adoption of LLMs among university students, with usage rates exceeding 45\% across multiple disciplines. This adoption has prompted significant concerns about academic integrity, with studies by \cite{a8} suggesting that traditional plagiarism detection tools are largely ineffective against LLM-generated content.
Several studies have systematically evaluated LLM performance in specific academic tasks. \cite{a9} found that LLMs achieved scores in the top percentile in standardized writing assessments, while \cite{b1} documented its ability to solve complex mathematical word problems with a high accuracy. In computer science education, \cite{b2} demonstrated that LLMs could successfully complete programming assignments across multiple languages, though performance declined significantly for tasks requiring complex algorithm design.
The impact on student learning outcomes has been more contentious. \cite{b3} reported improved conceptual understanding among students who used LLMs as learning aids and raised concerns about decreased retention of fundamental principles. Research by \cite{b5} suggests that the effectiveness of LLMs varies significantly based on the student's existing knowledge base and the complexity of the subject matter.
Institutional responses have varied widely. A comprehensive survey by \cite{b6} of 500 universities found approaches ranging from complete prohibition to active integration into curriculum design. 

\subsection{Existing Studies on AI in Engineering Education}
Several studies have focused on specific engineering domains. In electrical engineering, research by \cite{b9} evaluated LLM performance in circuit analysis, while mechanical engineering applications were extensively studied by \cite{b8}. However, these studies typically focused on specific problem types rather than complete course evaluation.
In control systems education--the topic of our paper--limited research exists on comprehensive LLM evaluation. Studies by \cite{c1} demonstrated successful LLM performance in control theory problems, research by \cite{c2} identified significant limitations in LLM's ability to generate and debug control system code, finding mixed results depending on the complexity of the implementation.
Prior work has begun exploring LLM performance in academic settings. \cite{c3} evaluated AI capabilities in mechanics coursework, while studies in computer science education by \cite{b3} and \cite{c5} have examined LLM integration in introductory programming and upper-level project work respectively. Our work builds on these foundations by providing a comprehensive assessment across all elements of an engineering course -- from theoretical problems to practical implementation and programming tasks.

\subsection{AE 353: Course Structure and Requirements}
\begin{figure}[htbp]
    \centering
    \includegraphics[width=0.45\textwidth]{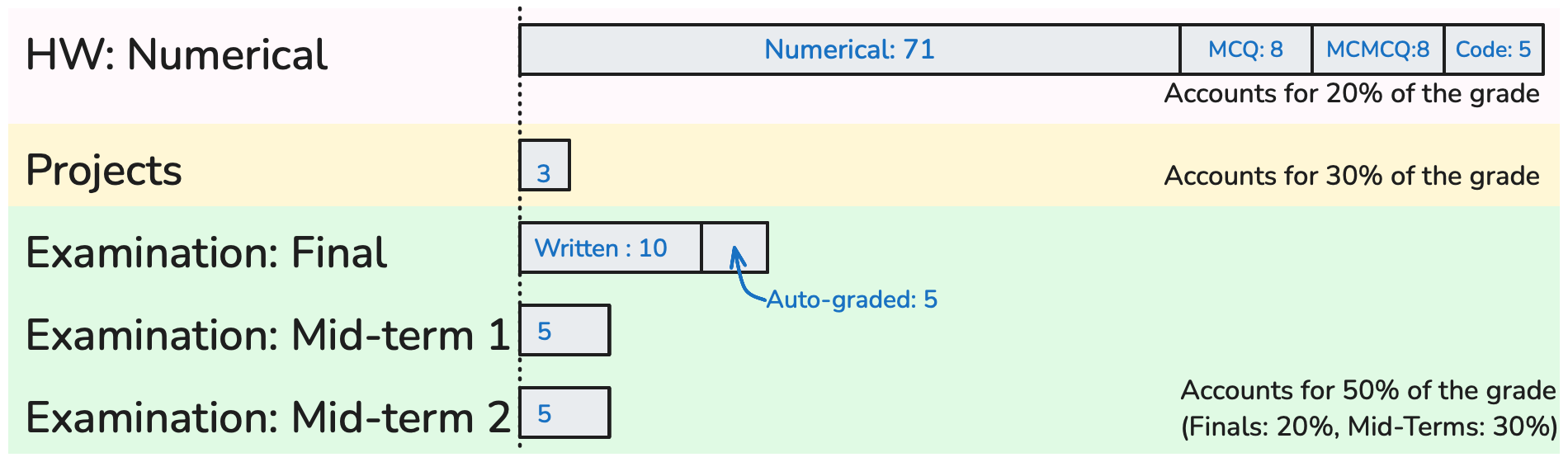} 
    \caption{Distribution of assessment types across AE 353.}
    \label{fig:dist} 
\end{figure}

The Aerospace Control Systems course (AE 353) 
 at UIUC serves as an introduction to control theory and its applications in aerospace systems. The course, mandatory for junior-level aerospace engineering students, integrates theoretical foundations with practical implementation through a diverse range of assessment types (see Fig. \ref{fig:dist} for the distribution of the assessment types). 

The course structure emphasizes progressive complexity in both theoretical understanding and practical application. Core concepts begin with fundamental control theory, before advancing to more sophisticated topics. This progression is reflected in the assessment structure, which comprises four distinct categories: homework assignments, midterm examinations, programming-intensive projects, and a comprehensive final examination.

Homework assignments, delivered through PrairieLearn, a platform that generates question variants with real-time feedback \citep{prairie1}, form the backbone of continuous assessment. These assignments test theoretical understanding with a combination of multiple-choice single correct questions (MCQ), multiple choice multi-correct questions (MCMCQ), numerical problems and basic Python programming tasks. The auto-graded nature of these assignments provides immediate feedback while maintaining consistent evaluation criteria.

The course includes two midterm examinations that assess the students' ability to demonstrate comprehensive problem-solving skills and theoretical understanding. Unlike homework assignments, midterms require detailed solution steps and allow partial credit, enabling a more nuanced evaluation of student understanding. The problems typically integrate multiple course concepts, requiring students to synthesize their knowledge across different topics.

Three substantial programming projects represent the course's practical implementation component. 
These proje-cts demand extensive Python coding, data analysis, and technical writing skills. Each project builds upon theoretical concepts covered in lectures, requiring students to implement control algorithms, analyze system behavior, and document their findings in detailed technical reports.

The final examination adopts a hybrid format, combining PrairieLearn-based auto-graded questions with traditional written problems. This structure allows for comprehensive assessment while maintaining efficiency in evaluation. The auto-graded portion consists of direct copies of homework problems (with PrairieLearn generating different variants of the question), while the written section consists primarily of midterm problems with minor modifications in parameters or problem statements.

\section{Methodology}
\subsection{Protocol Development}
\subsubsection{LLM Selection and Version}
In alignment with our objective to simulate minimal-effort student behavior, we selected the free version of ChatGPT (GPT-4) as our primary LLM. This choice reflects the model's widespread accessibility and its prominence among students. While more sophisticated LLMs exist, ChatGPT represents the most readily available tool for students seeking to complete coursework with minimal investment of time. 

\paragraph{Prompting Approaches}
Our investigation employed three distinct prompting approaches to ensure comprehensive assessment of LLM capabilities while maintaining the ``minimal effort" principle.

\begin{figure}[htbp]
    \centering
    \includegraphics[width=0.45\textwidth]{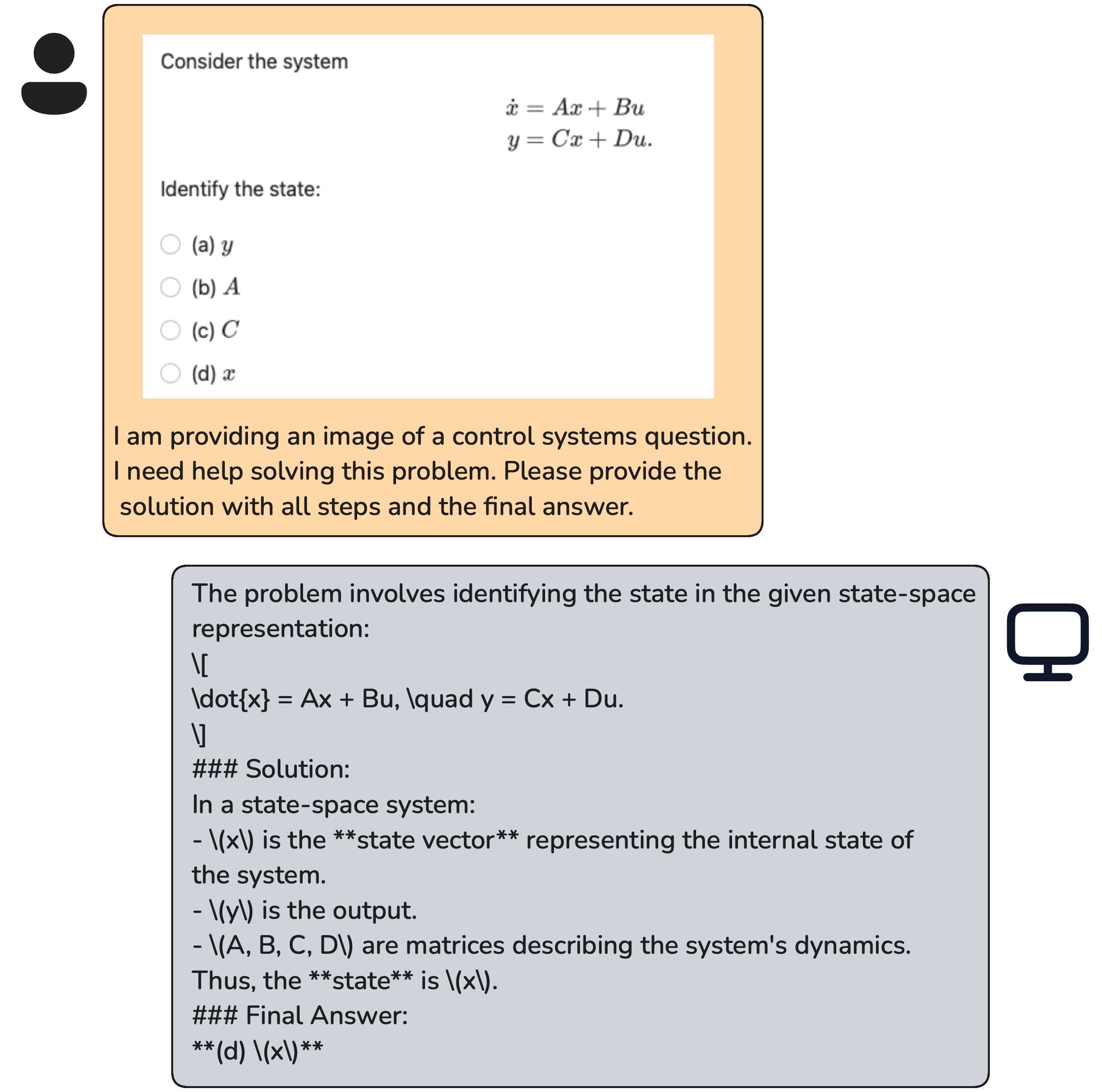}
    \caption{Example of image-based prompting.}
    \label{fig:image_prompt}
\end{figure}

\textit{Image-Based Prompting:} The first approach involved direct screenshot uploads of questions to ChatGPT, including mathematical equations, diagrams, and formatted text (Fig. \ref{fig:image_prompt}). This method most closely mimics a student's likely behavior of simply photographing or screenshotting assignment questions. While this method preserved the original formatting and visual context, it introduces challenges to the model's interpretation of mathematical notation and diagrams.

\begin{figure}[htbp]
    \centering
    \includegraphics[width=0.45\textwidth]{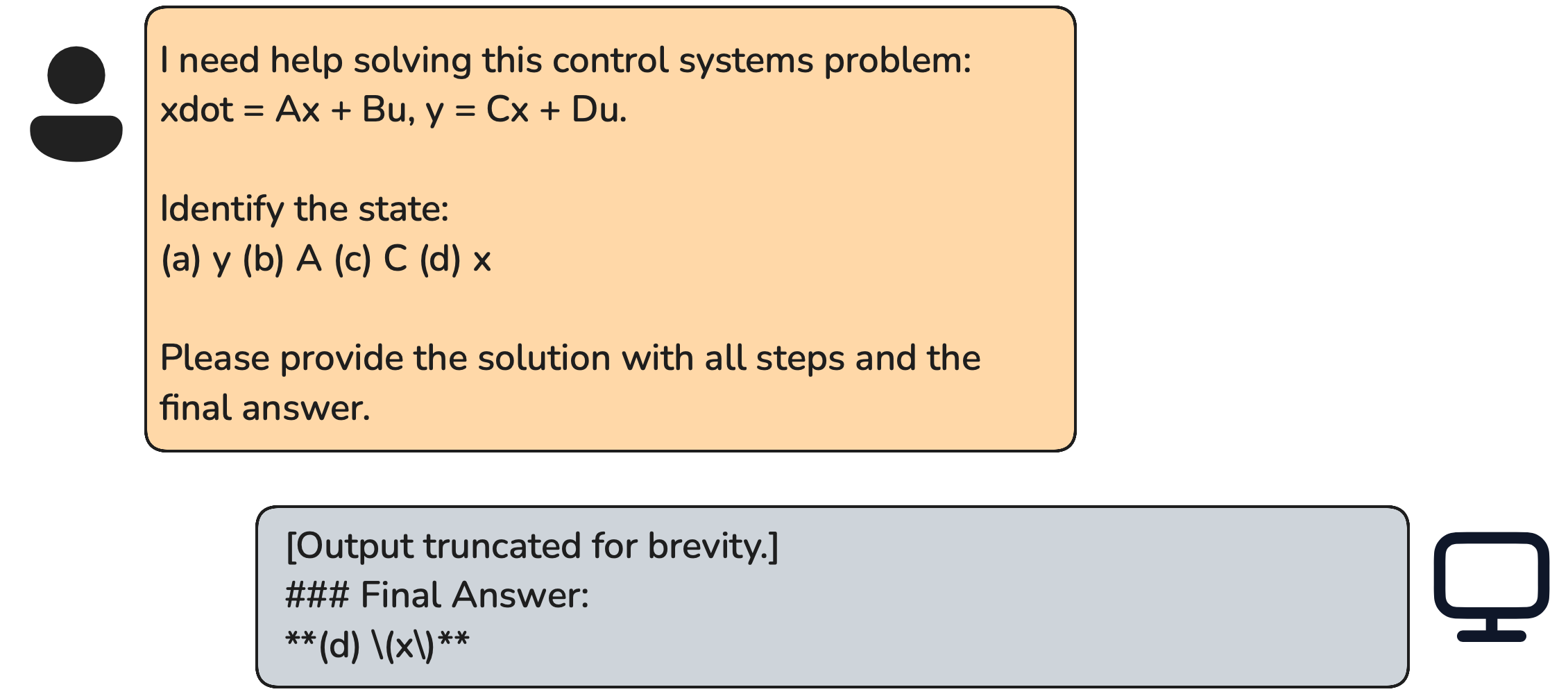}
    \caption{Example of simplified notation prompting.}
    \label{fig:simplified_math}
\end{figure}

\textit{Simplified Mathematical Notation:} Our second approach involved translating questions into simplified mathematical notation while maintaining essential mathematical relationships (Fig. \ref{fig:simplified_math}). This method tested the model's ability to process mathematical concepts without relying on formal mathematical typesetting.

\begin{figure}[htbp]
    \centering
    \includegraphics[width=0.45\textwidth]{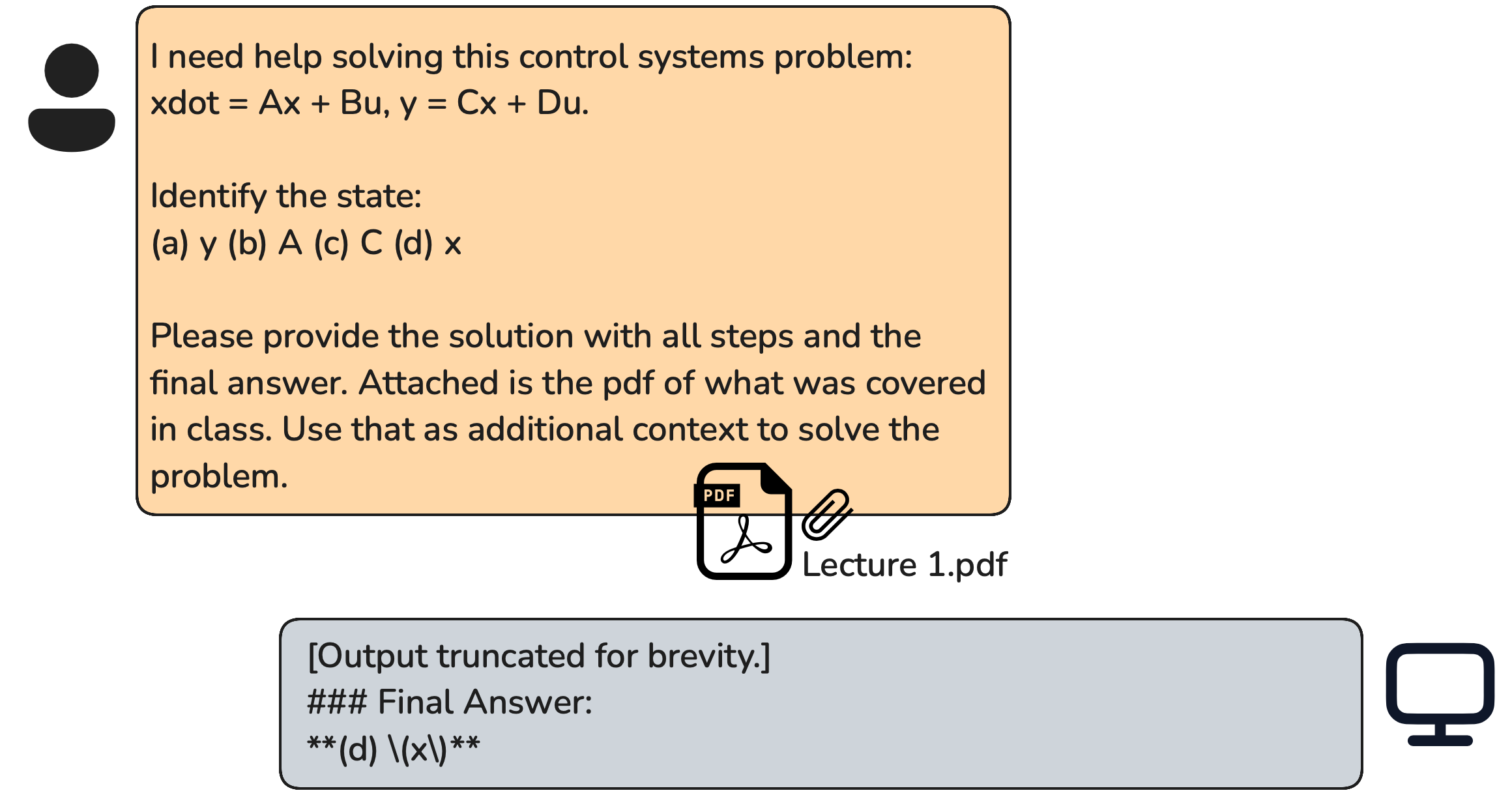}
    \caption{Example of context-enhanced prompting: Question presentation preceded by relevant lecture material.}
    \label{fig:context_enhanced}
\end{figure}

\textit{Context-Enhanced Prompting:} The final approach incorporated relevant basic lecture notes before presenting questions (Fig. \ref{fig:context_enhanced}). For each deliverable, we identified and included only the notes that covered the concepts being tested. These notes contained fundamental concepts from those specific lectures, rather than comprehensive textbook material. This approach tested whether minimal but relevant context could improve the LLM's performance while maintaining low-effort engagement.

Our testing methodology used both \textit{zero-shot} and \textit{multi-shot} approaches--essentially testing whether the model performed better when given multiple attempts with feedback versus a single attempt. While zero-shot prompting (presenting questions in isolation) remained consistent across all approaches, multi-shot implementations varied by assignment type. For PrairieLearn homeworks, we leveraged the immediate right/wrong feedback to refine subsequent attempts. Project submissions required sequential prompting--generating one section before moving to the next--while exam questions used structured follow-ups focused on mathematical verification. Note that context-enhanced prompting could not use zero-shot testing due to ChatGPT's token limitations, requiring us to split context and questions across messages.

\subsection{Translation Standards and Evaluation Methodology}
To ensure consistent evaluation, we developed standardized protocols for converting LLM outputs into gradable formats across all assessment types. For PrairieLearn homework assignments, translations were direct, using the model's first complete answer for multiple-choice, numerical responses, and code snippets. These responses required minimal formatting adjustments to meet platform requirements.
Mathematical derivations in examinations followed a rigorous translation process: textual expressions were converted to standard notation while preserving the exact sequence of steps and operations. We consistently implemented the first complete solution method when multiple approaches were provided. For programming projects, code outputs were preserved verbatim with only essential formatting changes for Python compatibility. Project reports underwent minimal editing, limited to satisfying formatting requirements specified in rubrics.
This approach maintained evaluation consistency while minimizing human intervention, simulating realistic student submission behavior across all assessment types. The translation process intentionally avoided any enhancement or correction of the LLM's responses, ensuring that evaluated content accurately reflected the model's raw capabilities.

%
%
\begin{figure*}[t]
    \centering
    \begin{minipage}[t]{0.22\textwidth}
        \centering
        \includegraphics[width=\textwidth]{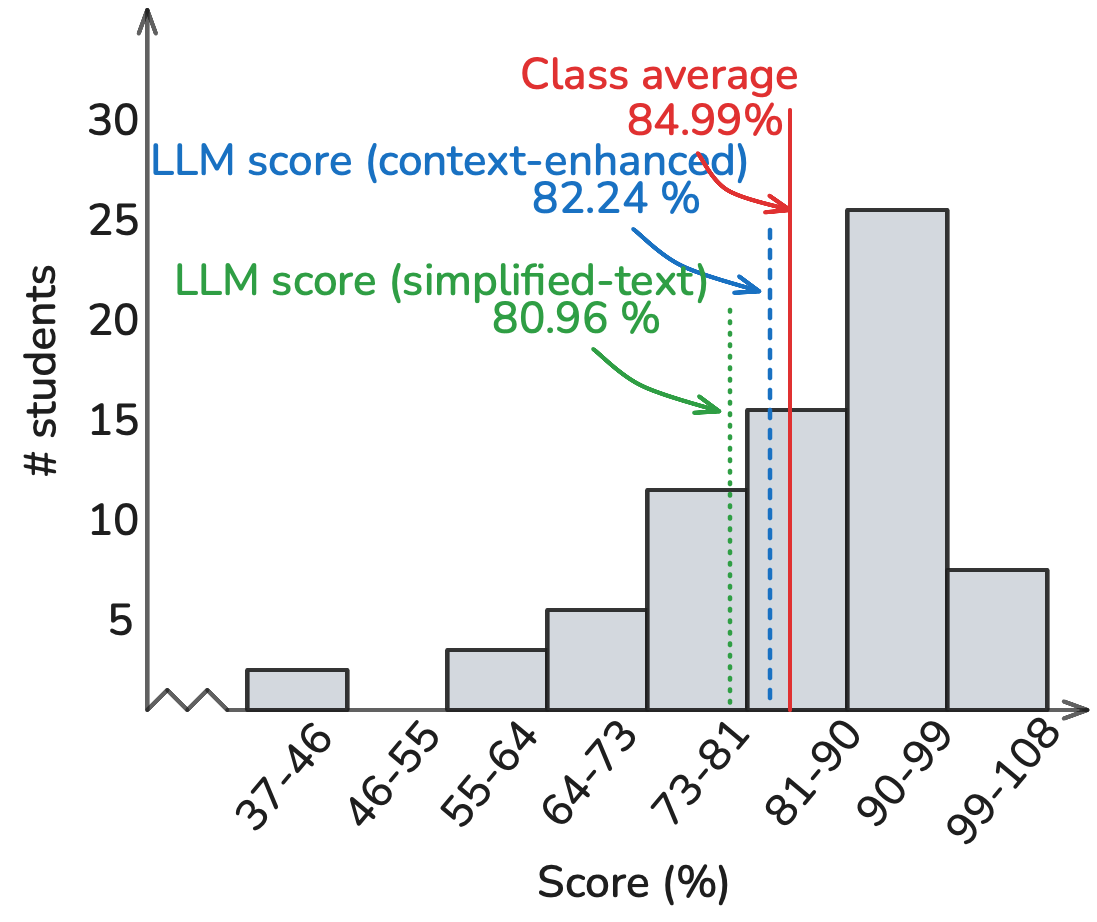}
        \vspace{1ex}
        \small (a) Overall performance
    \end{minipage}\hfill
    \begin{minipage}[t]{0.22\textwidth}
        \centering
        \includegraphics[width=\textwidth]{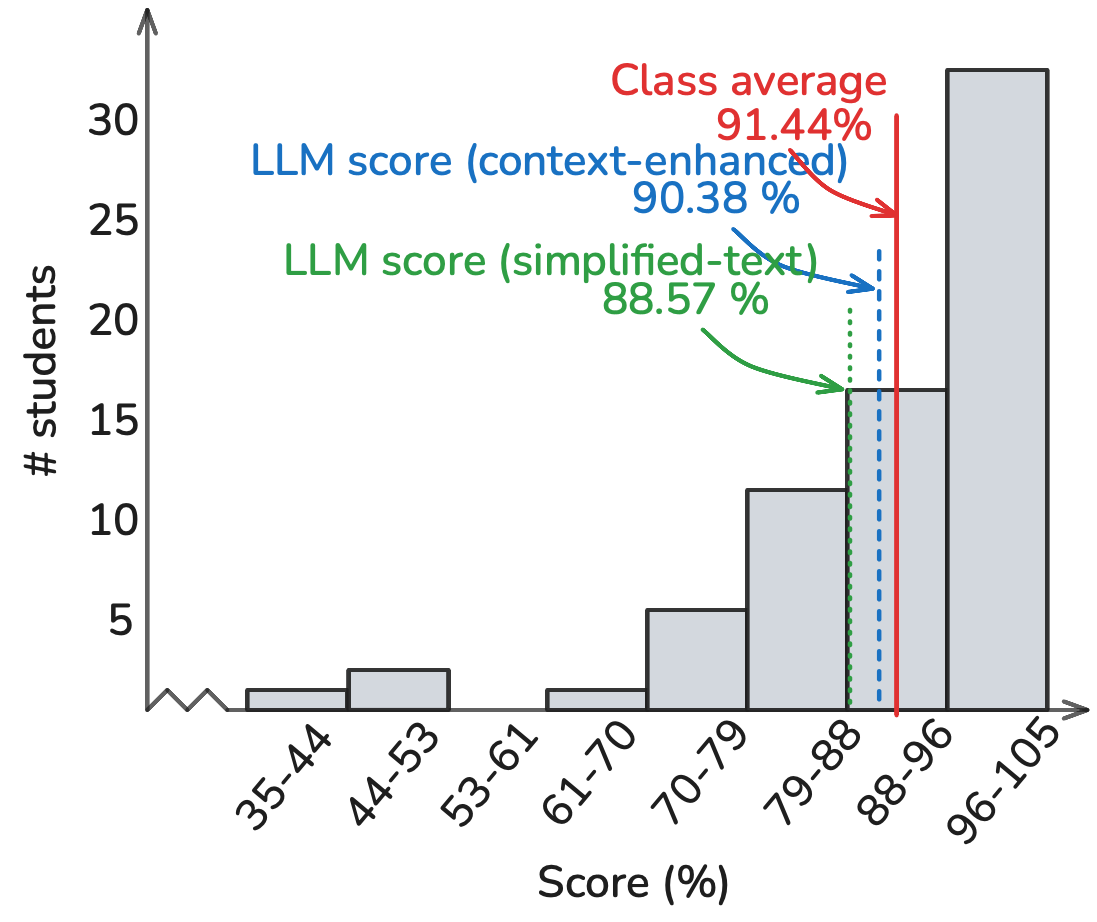}
        \vspace{1ex}
                \small (b) Homework performance
    \end{minipage}\hfill
    \begin{minipage}[t]{0.22\textwidth}
        \centering
        \includegraphics[width=\textwidth]{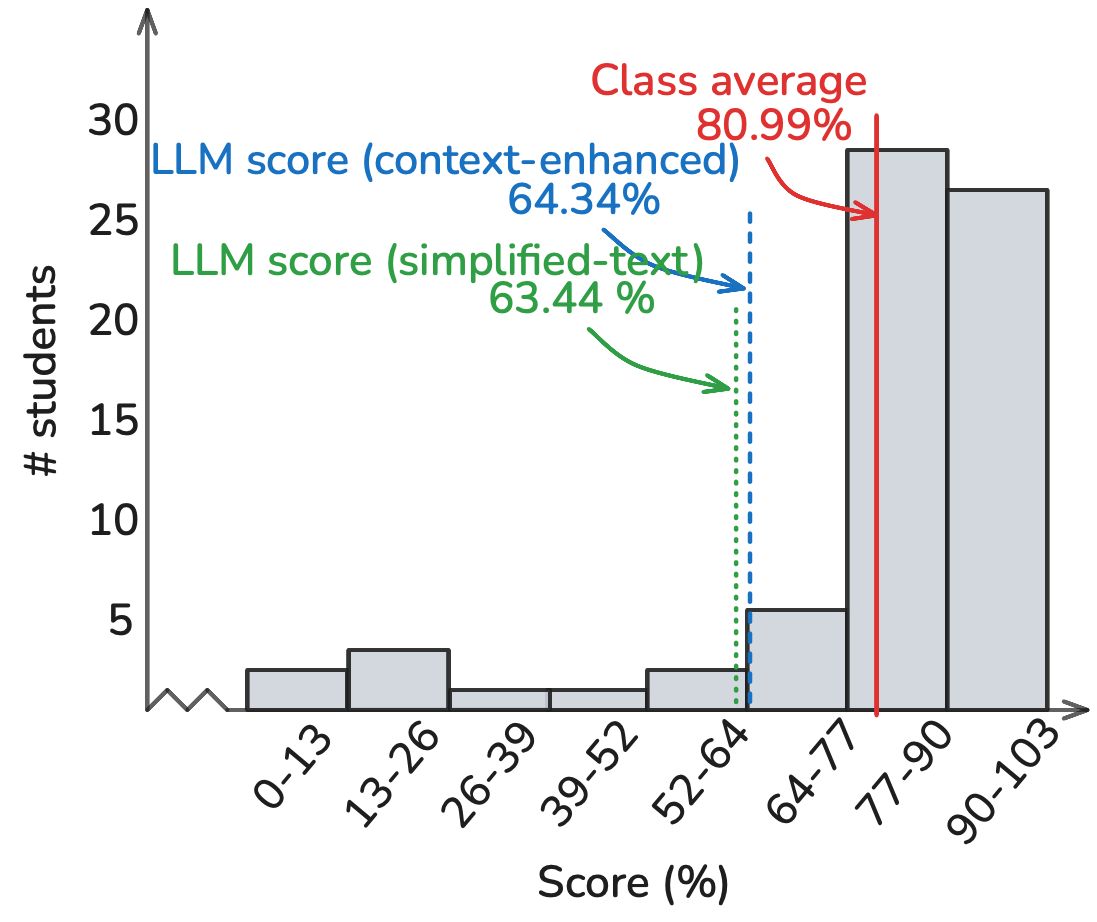}
        \vspace{1ex}
                \small (c) Project performance
    \end{minipage}\hfill
    \begin{minipage}[t]{0.22\textwidth}
        \centering
        \includegraphics[width=\textwidth]{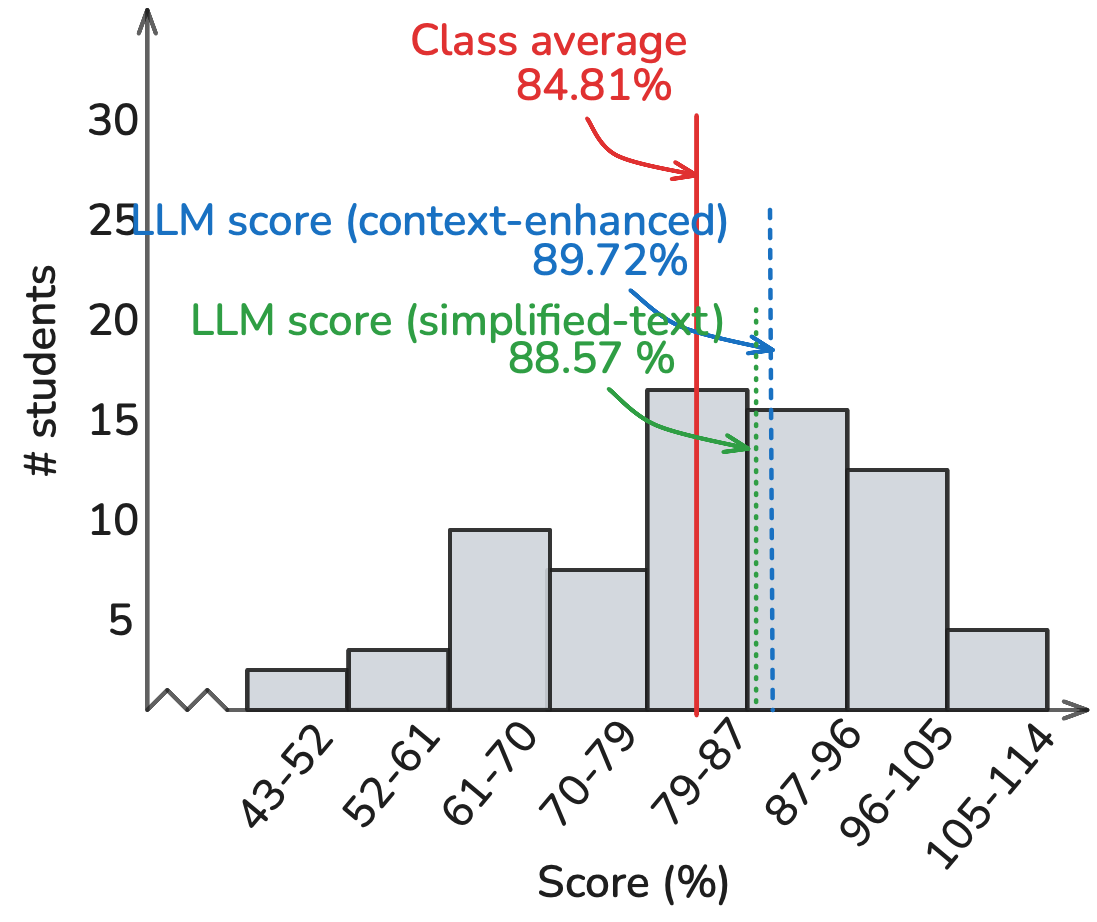}
        \vspace{1ex}
                \small (d) Exam performance
    \end{minipage}
    \caption{ 
Histograms depict the student cohort's score distribution. The LLM's performance is overlaid for context-enhanced prompting (blue, dashed) and simplified text (green, dotted) across different assignment types. Image-based prompting is excluded as projects were not graded using this method.}
    \label{fig:all_images}
\end{figure*}

\begin{table*}[htbp]
  \centering
  \renewcommand{\arraystretch}{1.2}
  \setlength{\tabcolsep}{8pt}
  \begin{threeparttable}
    \begin{tabular}{l|l|c|c|c|c|c}
      \toprule
      \multicolumn{2}{c|}{\textbf{Question Type}} & \multicolumn{2}{c|}{\textbf{Image Based}} & \multicolumn{2}{c|}{\textbf{Simplified Text}} & \textbf{Context-Enhanced} \\
      \cmidrule(lr){1-2} \cmidrule(lr){3-4} \cmidrule(lr){5-6} \cmidrule(lr){7-7}
      Category & Sub-type & Zero-shot (\%) & Multi-shot (\%) & Zero-shot (\%)  & Multi-shot (\%)  & Multi-shot (\%)  \\
      \midrule
      \multirow{4}{*}{\makecell[l]{HW}} 
        & MCQ        & 89.5 & 93.2 & 91.2 & 94.8 & 96.5 \\
        & MCMCQ      & 85.3 & 88.4 & 86.7 & 90.2 & 92.1 \\
        & Numerical  & 82.4 & 85.6 & 84.2 & 87.5 & 89.3 \\
        & Code-Based & 86.5 & 89.8 & 88.2 & 91.3 & 93.2 \\
      \cdashline{1-7}
      \multirow{2}{*}{Projects} 
        & Code   & - & - & 56.2 & 57.6 & 58.5 \\
        & Report & - & - & 62.8 & 64.9 & 65.8 \\
      \cdashline{1-7}
      \multirow{3}{*}{Exam} 
        & Mid-Term & 85.3 & 87.5 & 86.7 & 88.8 & 89.8 \\
        & Finals:Written & 83.0 & 84.1 & 83.8 & 84.6 & 86.5 \\
        & Finals:Auto-graded & 93.5 & 95.3 & 94.8 & 96.2 & 97.4 \\
      \bottomrule
    \end{tabular}
        
\caption{LLM performance across assessment types using various prompting methodologies.}  
    \label{tab:question_types}  
\end{threeparttable}
\end{table*}

\section{Results and Analysis}

\subsection{Performance Analysis}
\textit{Overall performance: }Table \ref{tab:question_types}\footnote{Note: Homework and project bonus points for early submission were excluded from the LLM performance evaluation.}
shows LLM performance across assessment types and prompting methods, based on three runs per question. While ChatGPT's exact wording varied between runs, these variations had minimal impact on scoring. Several key patterns emerge: First, context-enhanced prompting consistently outperforms other methodologies across all question types. Second, the progression from image-based to text-based inputs shows systematic improvement, particularly in questions involving mathematical notation. 
The LLM achieved an overall score of 82.24\% using context-enhanced prompting, compared to the class average of 84.99\% -- both corresponding to a `B' grade. This performance difference manifests distinctly across assessments: minimal in structured assignments but substantial in open-ended assessments.

\textit{Homework Performance:}
Homework performance reveals subtle patterns across question types. The LLM achieved 90.38\% against a class average of 91.44\%, with performance varying by question type. MCQs showed the highest success, followed by code-based questions, MCMCQ, and numerical problems. This hierarchy persisted across all prompting methodologies, though with varying gaps. The multi-shot approach proved particularly effective for MCQs. Analysis of the 92 homework questions reveals that performance degradation correlated strongly with question complexity: single-concept questions saw higher success rates compared to integration-heavy problems.

\textit{Examination Performance:}
Examination analysis provides critical insights into LLM capabilities under different assessment conditions. The model achieved 89.72\% overall compared to the class average of 84.81\%, but this aggregate masks important variations. Auto-graded components of the final (97.4\%) significantly outperformed written sections (86.5\%), with midterm performance (89.8\%) showing intermediate results. This pattern held consistent across prompting methodologies. The performance gap between written and auto-graded components suggests fundamental differences in the model's ability to handle structured versus open-ended problems. Since the final exam questions closely mirrored midterm content (with solutions previously provided to students), we tested providing these solutions as additional context to the LLM during final exam evaluation. Interestingly, this intervention produced negligible performance improvement, suggesting limitations in the model's ability to effectively leverage relevant solution patterns across similar problems.

\textit{Project Performance:}
Project evaluation exposed systematic limitations in LLM capabilities, with the most significant performance gap observed (64.34\% versus class average 80.99\%). The distinction between code implementation and report writing reveals specific challenges. Code submissions showed consistent patterns of failure in system integration, error handling, and optimization, while maintaining basic functional correctness. Report analysis indicates stronger performance in methodology description and result presentation but weaker performance in critical analysis and design justification. Neither image-based nor multi-shot approaches provided significant improvements in project performance, suggesting fundamental limitations rather than methodology-dependent constraints.

\subsection{Qualitative Analysis of LLM Responses}
The LLM's responses across assessment types revealed distinct patterns in its problem-solving approaches and conceptual understanding. In theoretical questions in projects and examinations, the model demonstrated sophisticated mathematical reasoning but often lacked the intuitive understanding that characterizes expert human responses. The model's explanations, while mathematically sound, frequently followed rigid, template-like structures that failed to adapt to the specific nuances of each problem.

In programming assignments, the LLM exhibited an interesting dichotomy: while capable of generating syntactically correct code, it often produced inefficient or overly complex solutions. The model's code implementations typically favored direct, brute-force approaches over elegant solutions that would demonstrate deeper understanding of control system principles. This feature was particularly evident in projects requiring the implementation of controllers, where the code would technically fulfill requirements but lack the optimization and robustness considerations that characterize high-quality student submissions.

Written explanations in project reports revealed several characteristic patterns of LLM behavior. Most notably, the responses demonstrated a tendency toward inappropriate sophistication, introducing advanced concepts in basic undergraduate-level analyses. While comprehensive in covering required topics, these responses often followed rigid, template-based structures: observation, technical analysis, and future recommendations, regardless of the specific project requirements. The model's inclination to include unnecessarily precise numerical values without proper justification and its liberal use of advanced technical jargon often masked a lack of fundamental understanding.

\begin{figure}[htbp]
    \centering
    \includegraphics[width=0.46\textwidth]{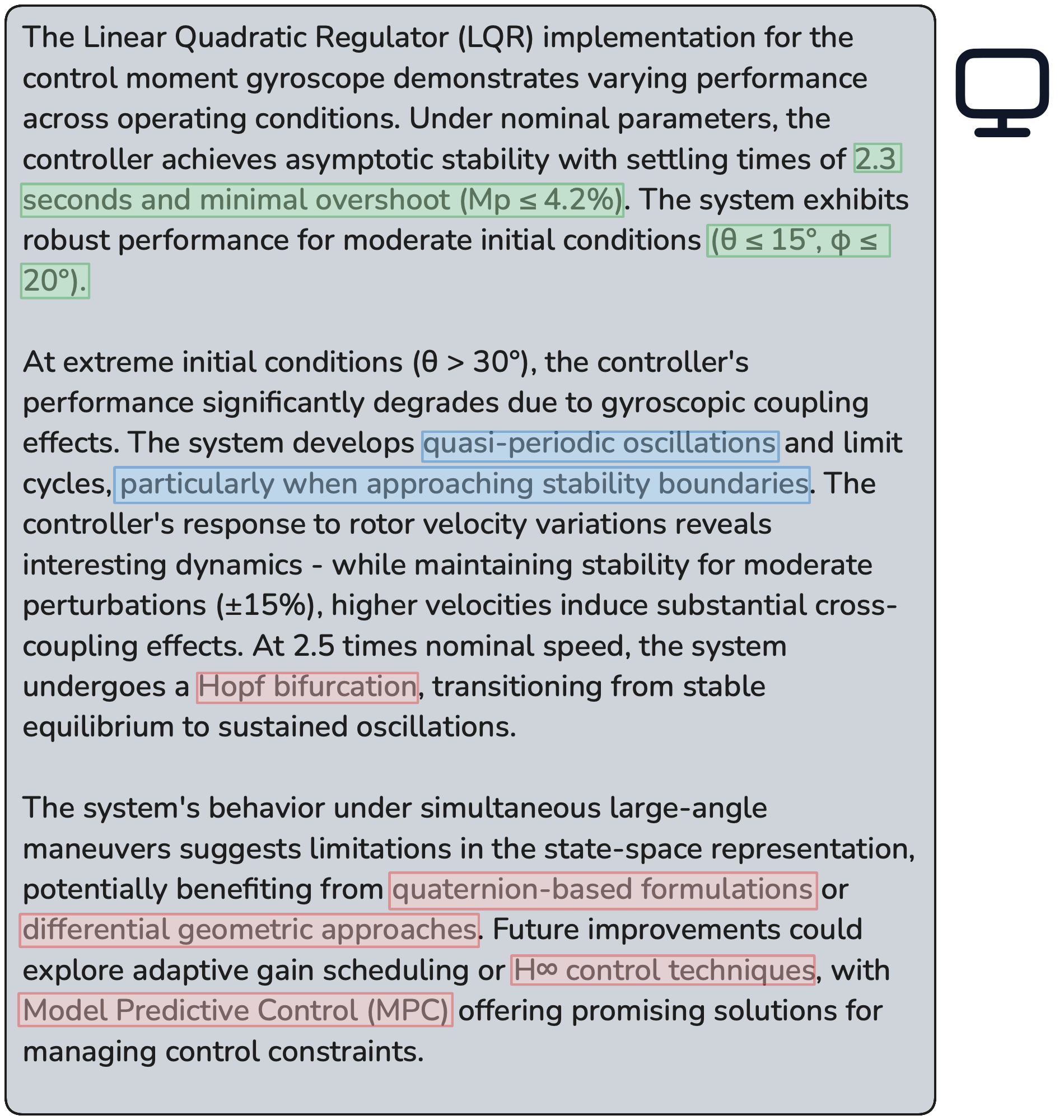} 
    \caption{Representative LLM response demonstrating characteristic patterns: hallucinated precise values (green), inappropriate technical jargon (blue), and advanced concepts beyond course scope (red).}
    \label{fig:example_image} 
\end{figure}

\section{Discussion}

Our findings demonstrate that LLMs, even with minimal effort approaches, can achieve passing grades in a rigorous undergraduate control systems course. The model's ability to earn a `B' grade raises questions about current assessment methods in engineering education. The performance disparity across different assessment types provides valuable insights into both the capabilities and limitations of current AI systems in technical education.
The strong performance in structured assignments (homework and auto-graded examinations) versus relatively weaker performance in open-ended projects suggests that the current assessment methods may inadvertently favor pattern recognition over deep understanding. 

\textit{Implications for Course Design: }
These results suggest a need for a fundamental reconsideration of assessment strategies in engineering education. Rather than solely focusing on how to prevent LLM use, we should consider how this technology might transform engineering education itself. 
Traditional evaluation methods, particularly those relying on structured problem-solving, may need to be redesigned to better differentiate between pattern recognition and genuine understanding. 

Several specific recommendations emerge from our analysis:
First, emphasis on integrated project work and open-ended design challenges could better assess students' ability to synthesize concepts and apply practical engineering judgment. The performance gap in project work suggests that such assignments remain effective at evaluating genuine understanding and technical capability.
Second, examination structures might benefit from focus on explaining reasoning and justifying engineering decisions rather than just arriving at correct answers. The LLM's difficulty with questions that require practical judgment indicates that such modifications could better assess human expertise.
Third, programming assignments should emphasize system integration, robustness, and optimization rather than basic implementation. The model's struggles with these aspects suggest that such requirements better evaluate genuine programming and engineering capability.

These recommendations, while addressing immediate assessment challenges, also point to a broader opportunity: by acknowledging and incorporating LLMs, we might free students from routine technical work to focus on higher-level engineering concepts and problem-solving skills. This shift could reinvigorate undergraduate education by reducing time spent on mechanical tasks in favor of conceptual understanding and practical engineering judgment.
%

\section{Conclusion}
This study provides concrete evidence that current LLMs can achieve passing grades in rigorous engineering coursework, even with minimal effort approaches. While this capability raises concerns about traditional assessment methods, it also highlights areas where human expertise remains distinct from AI capabilities. The significant performance gaps in project work and integrated problem-solving suggest that well-designed coursework can still effectively evaluate genuine engineering understanding and capability.
The findings suggest a need for evolution rather than revolution in engineering education. Moving forward, engineering education must adapt to acknowledge the reality of AI tools while continuing to develop and assess the unique capabilities that characterize effective human engineers.
These results provide a foundation for future research into AI's role in engineering education and suggest practical approaches for adapting course design to maintain educational effectiveness in an era of increasing AI capability. The challenge ahead lies not in preventing AI use, but in developing educational approaches that leverage these tools while continuing to cultivate genuine engineering expertise.

\section*{Acknowledgments}
This study was supported by the University of Illinois Urbana-Champaign’s Grainger College of Engineering through its Grants for the Advancement of Teaching in Engineering program.
The course's question bank represents several years of iterative development by the authors.
We gratefully acknowledge Grayson Schaer for building the project environments, and David Hanley and Pranay Thangeda for their significant contributions to the PrairieLearn question bank and lecture materials. We also thank the many others whose efforts helped make this research possible.

\bibliography{ifacconf.bib}

\end{document}